\documentclass[useAMS,usenatbib]{mn2e}

\title[Mechanism of radio emission]{The mechanism of coherent radio emission in some classes of pulsar}
\author[P. B. Jones]{P. B. Jones\thanks{E-mail:
p.jones1@physics.ox.ac.uk}  \\
University of Oxford, Department of Physics, Denys Wilkinson Building,\\
Keble Road, Oxford OX1 3RH, U.K.}

\begin{document}

\date{}

\pagerange{\pageref{firstpage}--\pageref{lastpage}} \pubyear{}

\maketitle

\label{firstpage}

\begin{abstract}

Radio luminosities have been estimated from published data for a well-defined homogeneous set of 29 normal pulsars.  The radio-frequency energies per unit charge in the primary accelerated particle beam are given for each pulsar and form a distribution more than two orders of magnitude in width.  The values found show that pulsars are extremely efficient generators of radiation below $1 -10$ GHz given the kinematic constraints which are obtained here in the cases of electron-positron and ion-proton plasmas.  Our conclusion is that only the ion-proton plasma source is consistent with the spectra of normal and millisecond pulsars and we list and describe the factors which together support this conclusion.

\end{abstract}


\section{Introduction}

Studies of pulsar radio luminosity in the past have been motivated by galactic population modelling or by an interest in the emission mechanism.  The work of Szary et al (2014) is a recent example of the latter. This present paper differs from earlier work in two respects: the luminosities are estimated for a well-defined and homogeneous set of modest size and, apart from the general shape of the spectra, the parameter we consider to be of prime importance is the radio-frequency energy $\epsilon$ emitted per unit charge in the  accelerated beam of primary particles. Owing to the invariably large negative spectral index, the low-frequency spectrum contains most of the luminosity.  Thus the selected pulsars each appear in the spectral compilation published by Malofeev et al (1994) and in the set found to display subpulse drift and selected by Weltevrede, Edwards \& Stappers (2006) purely on the basis of signal-to-noise ratio.

The details of the luminosity estimates are described in Section 2, and values of $\epsilon$ are given in Table 1. In Section 3, the emission volume is treated as a black box and we consider the extent to which kinematic constraints valid for photon production by free-particle beams can be applied to it.  On this basis, we compare the values of $\epsilon$ with expectations for the canonical electron-positron plasma and for an ion-proton plasma (see Jones 2012a,b; 2013a,b).

The wave number of the unstable mode to which particle longitudinal kinetic energy is transferred is dependent on number densities, rest masses and Lorentz factors.  Therefore, there will be considerable differences between predictions for the two types of plasma considered here.  We assume that mode growth proceeds through a stage of non-linearity to a turbulent distribution of charge-density and current fluctuations.  We further assume that development of the turbulence follows that in classical fluids and so has as its outstanding feature the transfer of energy to a spectrum of higher wavenumbers occurring on a very short time-scale. As a consequence, the mode frequency immediately prior to turbulence is to be associated with the lower frequencies in the radio-frequency spectrum and we do do not expect any significant radius-to-frequency mapping. The demonstration of an adequate growth rate for the ion-proton mode is of primary and obvious importance. The details of this have been published previously (Jones 2012a,b). That the mode frequency is apt has also been justified previously (Jones 2013a,b) and the details of both these questions will not be repeated here.  But they are considered further in relation to several important classes of pulsar in Section 4 and our conclusions are given in Section 5.

\section[]{Radio-frequency luminosities}

The spectral indices of pulsars are typically large and negative.  A large fraction of the radio-frequency power is at frequencies below $1$ GHz.  The pulsar luminosities we estimate are drawn from the survey of Malofeev et al (1994) who gave the spectra of 45 pulsars at these low frequencies.  In order to have a set as homogeneous as possible and with good signal-to-noise ratio, we required that each pulsar should also be one of the 70 listed by Weltevrede et al (2006) as exhibiting subpulse drift.  The set of 29 pulsars so defined can be characterized as normal (but not uninteresting) and is listed in Table 1.

The luminosity is defined by,
\begin{eqnarray}
L = d^{2}\delta\Omega\int^{\infty}_{0}S(\nu)d\nu,
\end{eqnarray}
in which $d$ is the distance, taken from the ATNF catalogue (Manchester et al 2005), and the integral is approximated by the two-index fit,
\begin{eqnarray}
\int^{\infty}_{0}S(\nu)d\nu \approx \frac{\alpha_{1}}{\alpha_{1}+1}\nu_{m}S_{m} + \frac{\alpha_{2} - \alpha_{1}}{(\alpha_{1} + 1)(\alpha_{2} + 1)}\nu_{b}S_{b},
\end{eqnarray}
noting that Malofeev et al give the flux densities per period $P$.  The indices $\alpha_{1,2}$ and frequencies $\nu_{m,b}$ are from Malofeev et al (Table 1) and the flux densities $S(\nu_{m})$ and $S(\nu_{b})$ have been obtained directly from their published  spectra.  Equation (2) assumes that, below the measured frequency interval, the flux density in and below the turn-over region is $S(\nu < \nu_{m}) = S(\nu_{m})$.

The solid angle $\delta\Omega$ is the most significant source of uncertainty in $L$.  It is defined as,
\begin{eqnarray}
\delta\Omega = 2\pi\sin\psi\delta\theta_{p},
\end{eqnarray}
where $\psi$ is the magnetic inclination and $\delta\theta_{p}$ is the increment in polar angle occupied by the beam and defined with respect to the spin axis ${\bf \Omega}$.  The solid angle is estimated for each of the 29 pulsars listed in Table 1 using the widths $W_{50}$ listed in the ATNF catalogue (Manchester et al 2005) and by making the following assumptions, in which averages refer to the group of 70 Weltevrede et al pulsars.

(i)		The beam profile is (arbitrarily) circular and its transit across the line of sight is represented by an arc subtending an angle $\delta\theta = 2\pi\sin\psi W_{50}/P$ on the surface of the unit sphere.

(ii)	For random intersection of a circular or semi-circular profile, $\delta\theta_{p} = (4/\pi)\delta\theta$.

(iii)	The magnetic co-latitude of flux lines bounding the open magnetosphere at any altitude above the polar cap considered here can be expressed as $\delta\theta = \delta\theta_{0}(P_{0}/P)^{1/2}$, where $P_{0} = 1$ s.

(iv)	The sources of emission in the Weltevrede et al group of 70 pulsars are at a fixed altitude $\eta$, in units of the neutron-star radius $R$.

The distribution of $\delta\theta_{0}$ should have little variance compared with those of $\sin\psi$ and $W_{50}$ so that, by averaging over the Weltevrede group of 70 pulsars, it is a fair approximation to express its average as,
\begin{eqnarray}
\langle\delta\theta_{0}\rangle = \left\langle\frac{2\pi W_{50}}{(PP_{0})^{1/2}}\right\rangle\frac{1}{\langle\csc\psi\rangle}.
\end{eqnarray}
For any individual pulsar, an estimate of the magnetic inclination is then given by,
\begin{eqnarray}
\sin\psi = \frac{\langle\delta\theta_{0}\rangle(PP_{0})^{1/2}}{2\pi W_{50}}.
\end{eqnarray}
In this expression, $\langle\csc\psi\rangle$ is unknown.  An initial assumption of an isotropic distribution, $\langle\csc\psi\rangle = \pi/2$, gives values of $\sin\psi$ that are quite inconsistent with that hypothesis in that there are large excesses of unphysical values and of small $\sin\psi$.  The unphysical values arise, given assumptions (i) - (iv), because cases of partial intersection of the beam profile occur and are associated with too small a value of $W_{50}$.
Therefore, assuming that (i) - (iv) are sound, it is possible to adjust the unknown quantity $\langle\csc\psi\rangle$ so as to reduce the number of unphysical values to a level estimated to be consistent with the distribution in the interval $0.5 < \sin\psi < 1$. This procedure appears worthwhile because values of $\sin\psi$ can vary by as much as an order of magnitude between pulsars.  On this basis, a value $\langle\csc\psi\rangle = 3.5$ has been adopted, giving $\langle\delta\theta_{0}\rangle = 0.051$ radians and the values of $\sin\psi$ in Table 1, the remaining unphysical values being denoted by $\sin\psi = 1.0$.  It also gives, through assumptions (ii) and (iii), the value of $\delta\theta_{p}$ for equation (3). It has to be emphasized that this distribution, which is strongly peaked at small $\sin\psi$, is specific to the Weltevrede et al group of 70 pulsars and there is no suggestion that it is of any universal relevance.

\begin{table}
\caption{Estimated radio luminosities $L$ are given for a set of 29 pulsars.  Assuming Goldreich-Julian current densities at the polar cap, the radio-frequency energy $\epsilon$ per unit charge in the primary accelerated particle beam has been calculated.  The periods and distances $d$ are from the ATNF catalogue (Manchester et al 2005) as are the values of the pulse widths $W_{50}$ used to calculate the magnetic inclination angle $\psi$ on the basis of the polar-cap model assumptions (i) - (iv) of Section 2.  The integrals over flux density are given by equation (2), with parameters from Malofeev et al (1994).}

\begin{tabular}{@{}lrccrrc@{}}
\hline
Pulsar   &  $P$  &  $d$  &  $\sin\psi$ & $\int^{\infty}_{0}Sd\nu$ & $L$ & $\epsilon$  \\
\hline
         &   s   &  kpc  &           &  mJy GHz & erg s$^{-1}$  &  GeV  \\
\hline
B0138+59  & 1.223 &  2.3 & 0.259 & 44  & 2.1E27  &  3.3      \\
B0809+74  & 1.292 &  0.43& 0.224 & 188 & 2.7E26  &  0.71     \\
B0818-13  & 1.238 &  1.9 & 0.441 &  32 & 1.8E27  &  1.2      \\
B0919+06  & 0.431 & 1.1 & 0.511 & 124 & 4.5E27 & 0.29  \\
B1508+55  & 0.740 & 2.1 & 0.643 & 182 & 2.3E28 & 4.7 \\
B1604-00  & 0.422 & 0.59 & 0.426 & 64 & 5.7E26 & 0.21 \\
B1845-01  & 0.659 & 4.4 & 0.391 & 67 & 2.4E28  & 4.1  \\
B1911-04  & 0.826 & 3.2 & 0.984 & 62 & 2.7E28 & 7.3  \\
B2044+15  & 1.138 & 2.6 & 0.908 & 15 & 3.3E27 & 6.8 \\
B2154+40  & 1.525 & 2.9 & 0.261 & 32 & 2.2E27 & 1.6 \\
B2319+60  & 2.256 & 2.7 & 0.094 & 29 & 5.1E26 & 0.48 \\
B2351+61  & 0.945 & 3.3 & 0.764 & 30 & 9.8E27 & 1.6 \\
B0031-07  & 0.943 & 1.0 & 0.140 & 55 & 3.1E26 & 0.32  \\
B0320+39  & 3.032 & 1.5 & 0.333 & 56 & 9.3E26 & 4.5  \\
B0329+54  & 0.715 & 1.0 & (1.0) & 956 & 4.4E28 & 14  \\
B0525+21  & 3.746 & 2.3 & 0.085 & 82 & 7.4E26 & 0.62 \\
B0628-28  & 1.244 & 0.32 & 0.157 & 198 & 1.1E26 & 0.04 \\
B0823+26  & 0.531  & 0.32 & (1.0) & 134 & 7.3E26 & 0.16 \\
B0834+06  & 1.274 & 0.72 & 0.385 & 177 & 1.2E27 & 0.48  \\
B1133+16  & 1.188 & 0.35 & 0.281 & 370 & 4.5E26 & 0.22  \\
B1237+25  & 1.382 & 0.84 & 0.188 & 309 & 1.4E27 & 1.7  \\
B1642-03  & 0.388 & 2.9 & (1.0) & 430 & 2.2E29 & 29 \\
B1929+10  & 0.227 & 0.31 & 0.527 & 198 & 8.2E26 & 0.06  \\
B1933+16  & 0.359 & 3.7 & 0.545 & 189 & 9.1E28 & 5.7  \\
B1944+17  & 0.441 & 0.30 & 0.402 & 107 & 2.3E26 & 0.33 \\ 
B2016+28  & 0.558 & 0.98 & 0.411 &124 & 2.5E27 & 2.0  \\
B2021+51  & 0.529 & 1.8 & 0.801 & 130 & 1.8E28 & 2.8  \\
B2045-16  & 1.962 & 0.95 & 0.136 & 135 & 4.6E26 & 0.28  \\
B2111+46  & 1.015 & 4.0 & 0.256 & 310 & 4.9E28 & 43 \\

\hline

\end{tabular}

\end{table}

Table 1 gives the estimated luminosities for the set of 29 pulsars selected.  Following Harding \& Muslimov (2001), a polar-cap radius $u_{0}$ given by,
\begin{eqnarray}
u_{0} = \left(\frac{2\pi R^{3}}{cPf(1)}\right)^{1/2},
\end{eqnarray}
is assumed, with a neutron-star mass of $1.4M_{\odot}$ and radius $R = 1.2\times10^{6}$ cm, for which $f(1) = 1.368$.  The radio-frequency energy emitted per unit primary charge is $\epsilon = L/f_{GJ}(0)$, where $f_{GJ}(0)$ is the Goldreich-Julian flux of unit charges from an area of $\pi u_{0}^{2}/2$ assuming $\psi = 0$ and calculated using the ATNF values of magnetic flux density. (The flux $f_{GJ}$ is a function of the total magnetic flux crossing the polar cap, not the surface magnetic flux density, and for present purposes we do not assume that the $\cos\psi$ dependence of the Goldreich-Julian density is valid in the vicinity of $\psi = \pi/2$. It is also assumed that one half of the open flux lines intersecting the polar cap area do not have the favourable direction of curvature at larger values of $\eta$ that would allow polar-cap acceleration.) This is the quantity we regard as most relevant to the emission process.  The values listed in Table 1 vary by more than two orders of magnitude and some, at first sight, seem surprisingly large. The average is $\bar{\epsilon} = 4.7$ Gev, and with the removal of B0329+54, B1642-03 and B2111+46 would be reduced to $1.9$ GeV. A caustic in the variable $\theta$ might be a possible explanation for this small number of very radio-bright pulsars.

In general, objections that our approximation for $\delta\Omega$ under-estimates luminosities would be unlikely to alter the conclusions arrived at in this paper.
The luminosities in Table 1 are typically between one and two orders of magnitude smaller than the mean value $L \approx 10^{29}$ erg s$^{-1}$ obtained by Szary et al (2014) for 1436 pulsars from ATNF values of the flux density at 1.4 GHz.  Their distribution of $L$ has a large variance, extending over about three orders of magnitude, and they find that $L$ is independent of position in the $P-\dot{P}  $ plane.

\section[]{Kinematics of the emission process}

We shall assume that the basis of the emission process is the formation of a plasma within the tube of open magnetic flux lines, principally above the acceleration region but typically at altitudes $\eta < 2$, which moves outward to the light cylinder.  In the case of an electron-positron plasma, its initial velocity distribution may change slowly so as to maintain the charge density needed locally to remove any electric field component ${\bf E}_{\parallel}$ arising from motion along curved flux lines.  (Here, the subscripts perpendicular and parallel refer to the local magnetic flux density ${\bf B}$.)  In the course of this motion, a fraction of the longitudinal kinetic energy of the plasma particles is transferred to an unstable Langmuir mode, which may be quasi-longitudinal (see Asseo, Pelletier \& Sol 1990), and eventually to the radio-frequency spectrum of the radiation field.

For kinematic purposes, the section of the flux tube from plasma formation to radio-frequency emission can be considered as a black box. The plasma in the emission region is a one-dimensional freely moving system in the frame of reference corotating with the neutron star apart from external influences which include the Coriolis and flux-line curvature accelerations, and the longitudinal force arising from any small ${\bf E}_{\parallel}$ which may be present.  Flux-line curvature is the source of coherent curvature radiation, but for reasons to be given in Section 4, it is not considered further here.  However, the large values of $\epsilon$ given in Table 1 show that ${\bf E}_{\parallel}$ can be very significant.

The kinematics of radio-frequency generation is independent of any assumption about the interior processes of the black box.  However, the mode growth rate, at least in the case of an ion-proton plasma, is large, leading to non-linearity and the eventual development of a turbulent system of charge and current-density fluctuations.  Given the presence of transverse spatial gradients, particularly in the velocity distribution, and of finite transverse dimensions, the form of the turbulence is unlikely to have any simplicity of structure and there would appear to be no reason to assume it unable to couple with both modes of the radiation field.  A common feature of turbulent systems is the transfer of energy to larger wave-numbers on rapid time-scales.  If this is so, we should expect that the complete radio spectrum is formed within a small region so that radius-to-frequency mapping will not be apparent.

\subsection{Electron-positron plasma}

From the ATNF catalogue fields, it appears that the magnetic flux density in the formation zone for the pulsars listed in Table 1 is in the classical region of single-photon magnetic conversion (Erber 1966: see also Harding \& Lai 2006).  Its transition rate is determined by the parameter $\chi = k_{\perp}B/B_{c}$, where $k$ is the photon momentum in units of $m_{e}c$ and $B_{c} = 4.41\times10^{13}$ G is the critical field.  For the length scales in the present case, pair formation by conversion requires $\chi \sim 0.1$.  The form of the electron-positron spectrum can be found most easily by a Lorentz transformation of the photon parallel with ${\bf B}$ from the neutron-star frame to that in which $k_{\parallel} = 0$. Then the pair has total energy $k_{\perp}$, and its centre-of-mass velocity and Lorentz factor are $k_{\parallel}/k$ and $k/k_{\perp}$, respectively, relative to the neutron-star frame.  In the classical region, values of $k_{\perp}$ are substantially above the threshold $k_{\perp} = 2$ so that electron and positron Landau quantum numbers are not small. A high pair multiplicity $N_{\pm}$ per primary accelerated electron or positron ensures that ${\bf E}_{\parallel} = 0$ in the formation zone.  Owing to the charge symmetry of the conversion process, electrons and positrons separately have identical energy distributions with centre-of-mass energies $\tilde{m}$ per primary particle with, as a result of there being a distribution of values of $k_{\parallel}$, $\tilde{m} > \langle k_{\perp}\rangle N_{\pm}/2$ in terms of the average of $k_{\perp}$.  The three centres of mass, for electrons and positrons separately, and for the system of electrons and positrons combined, initially have identical velocities in the neutron-star frame.

The total charge density $\rho$ of primary beam and secondary pairs adjusts during outward flow so as to remove any ${\bf E}_{\parallel}$ component that arises from flux-line curvature. Thus $\rho = \rho_{GJ}$, where $\rho_{GJ}$ is the  Goldreich-Julian corotational charge density . The charge density of the primary beam is approximately equal to $\rho_{GJ}$ and the secondary electrons and positrons are initially neutral. Let the electron and positron centres of mass have Lorentz factors $\gamma_{c-,c+}$, respectively, the combined system having Lorentz factor $\gamma_{c}$.  Adjustment of the secondary electron and positron charge density occurs through a small change in relative velocity. Assume that a momentum increment $\pm\delta p$ is imparted to each particle.  Then by direct computation, we find that the centre-of-mass energy $\tilde{m}$ of either electrons or positrons is changed only to order $(\delta p)^{2}$, whilst the centre-of-mass Lorentz factors are changed to order $\delta p$.  For motion along a flux-line of increasing (positive) Goldreich-Julian charge density, the initial state, $\gamma_{c-} = \gamma_{c+} = \gamma_{c}$ is changed to a final state $\gamma_{c+} < \gamma_{c} < \gamma_{c-}$. (Alternatively, in the combined centre-of-mass system, we can see that in this process no momentum is imparted to the pairs, but both electrons and positrons gain energy.)

The initial state has identical electron and positron velocity distributions and thus has no charge or current density fluctuations: it is the ground-state of the system.  Hence the maximum radio-frequency energy that can be produced in the absence of any external forces acting on the plasma is determined by the energy difference between the ground-state and the momentum-displaced state.  From conservation of energy and of momentum parallel with ${\bf B}$ it is,
\begin{eqnarray}
W = \frac{(\gamma_{c-} - \gamma_{c+})^{2}}{\gamma_{c-} + \gamma_{c+}} \tilde{m}c^{2},
\end{eqnarray}
in the neutron-star frame, per unit electron or positron charge in the primary beam.

For the purposes of calculation, we require a specific form for the energy spectra of electrons and positrons in the neutron-star frame.  The number densities $n_{+,-}$ per unit interval of the Lorentz factors $\gamma_{+,-}$ are
uniform and initially equal, $\gamma_{1} \leq \gamma_{+,-} \leq \gamma_{2}$, where $\gamma_{2} \gg \gamma_{1} \gg 1$.  The momentum displacement changes the intervals to $\gamma_{1,2} + \Delta$ and $\gamma_{1,2} - \Delta$, and the centre-of-mass Lorentz factors so that $\gamma_{c-} - \gamma_{c+} = 2\Delta$. 
The condition for the secondary electron-positron current density to be zero defines the difference $n_{+} - n_{-}$ and is,
\begin{eqnarray}
\int^{\gamma_{2}+\Delta}_{\gamma_{1}+\Delta}n_{-}\left(\frac{\gamma_{-}^{2} -1}
{\gamma_{-}^{2}}\right)^{1/2}d\gamma_{-} =   \nonumber   \\
\int^{\gamma_{2}-\Delta}_{\gamma_{1}
-\Delta}n_{+}\left(\frac{\gamma_{+}^{2}-1}{\gamma_{+}^{2}}\right)^{1/2}d\gamma_{+}.
\end{eqnarray}
In terms of the number densities, the charge excess is,
\begin{eqnarray}
\delta \rho = {\rm e}(n_{+} - n_{-})(\gamma_{2} - \gamma_{1})
\end{eqnarray}
and the Goldreich-Julian charge density,
\begin{eqnarray}
\rho_{GJ} \approx \frac{{\rm e}(n_{+} + n_{-})(\gamma_{2} - \gamma_{1})}{2N_{\pm}}.
\end{eqnarray}
An approximation for $\Delta$ then follows from equation (8),
\begin{eqnarray}
\Delta = \frac{\gamma_{1}^{2}\gamma_{2}}{N_{\pm}}\frac{\delta \rho}{\rho_{GJ}},
\end{eqnarray}
which is valid in the limit $\gamma_{2} \gg \gamma_{1} \gg 1$.  This demonstrates the fact that charge-density adjustment depends almost entirely on electrons and positrons near the low energy limits of their spectra in the neutron-star frame.  In reality, the limit assumed, $\gamma_{1} \gg 1$ is unlikely to be satisfied and $\gamma_{1}$ will usually be close to unity.

The expected order of magnitude of the charge-density adjustment is $\delta \rho/\rho_{GJ} \sim 10^{-2}$ occurring over a length of $10^{6}$ cm. Equation (11) shows that $\Delta$ is small compared with unity for a wide range of values of $\gamma_{2}$ and $N_{\pm}$.  The implication is that the radio-frequency energy limit given by equation (7) is very small compared with the values of $\epsilon$ listed in Table 1.  But this in itself does not necessarily exclude an electron-positron plasma in these pulsars.

The effect of radio-frequency emission by whatever mechanism is to move the electron and positron centre-of-mass Lorentz factors $\gamma_{c-,c+}$ towards
$\gamma_{c}$.  The requirement that the total charge density $\rho$ of the primary beam and of the electron-positron plasma should be equal to $\rho_{GJ}$ has the opposite effect and increases the energy of the plasma above its ground state.  Without specifying the precise mechanism, we could assume that the energy transfer in the continuous charge-density adjustment then supports the radio-frequency emission, thus circumventing the limit given by equation (7).

A separate problem arising from small values of $\Delta$ is the question of whether the relativistic Penrose condition (see Buschauer \& Benford 1977)  for growth of an unstable mode can be satisfied.  This requires that the plasma particle velocity distribution has two separate maxima. Many years ago, Cheng \& Ruderman (1977) recognized that flux-line curvature would, in principle, allow this, but small values of $\Delta$ make it problematic.

Owing to a lack of knowledge, we have assumed so far that the electron and positron number densities are uniform functions of $\gamma_{+,-}$.  But it is just possible that the combined distribution, with centre-of-mass Lorentz factor $\gamma_{c}$ may itself have two maxima, satisfying the relativistic Penrose condition, independently of any electrostatic separation which is present.  The most basic approximation would be that the number of secondary electrons plus positrons per primary electron or positron is $N_{\pm}(\delta(\gamma - \gamma_{1}) + \delta(\gamma - \gamma_{2}))$ with $\gamma_{1} \ll \gamma_{2}$.  The maximum radio-frequency energy is then,
\begin{eqnarray}
W = N_{\pm}mc^{2}\frac{(\gamma_{1} - \gamma_{2})^{2}}{\gamma_{1} + \gamma_{2}},
\end{eqnarray}
per unit electron or positron charge in the primary beam. This could exceed that given by equation (7) and we shall consider it briefly in Section 4.

\subsection{Ion-proton plasma}

The kinematics of ion-proton plasmas is more simple. The protons on any flux line have a $\delta$-function velocity distribution.  Ions are slightly more complex because there can be a distribution of ion charges $\tilde{Z}$ for a given atomic number $Z$; also the possibility of different mass numbers $A$.  Nonetheless, a single $\delta$-function approximation for the ions is satisfactory for present purposes.  The physical system differs from the electron-positron plasma case in that the ions and protons are the primary beam particles and obviously satisfy the Penrose condition as a consequence of their different charge-to-mass ratios.
We refer to Jones (2012a,b) for a more complete account of the formation and properties of this system.

In the neutron-star frame, the energy associated with unit positive charge in the beam is,
\begin{eqnarray}
\frac{m_{p}c^{2}}{1 + y} \left(x\gamma_{I} + y\gamma_{p}\right),
\end{eqnarray}
in terms of the Lorentz factors $\gamma_{I}$ and $\gamma_{p}$, where $x = A/\tilde{Z}$, $y$ is the mean number of free protons in the beam per unit positive charge of the ions, and $m_{p}$ is the proton mass.  Proceeding from conservation of energy and of momentum parallel with ${\bf B}$, as in Section 3.1, the maximum energy that can be transferred to radio-frequency photons is, assuming $\gamma_{I,p} \gg 1$
\begin{eqnarray}
W = \frac{xy(\gamma_{p} - \gamma_{I})^{2}}
{(1 + y)(x\gamma_{p} + y\gamma_{I})}m_{p}c^{2},
\end{eqnarray}
in the neutron-star frame, per unit positive charge in the beam.  The ground-state of the system is that in which the Lorentz factors are $\gamma_{I} = \gamma_{p} = \gamma_{c}$, where $\gamma_{c}$ is that for the centre of mass.
The ratio $x$ is likely to satisfy $2 < x < 4$, and $y$ is a function of the mass number of nuclei at the surface of the particular neutron star and of the acceleration potential difference on the bundle of flux lines concerned.  We refer to Jones (2012) for more complete details but note that $y = 2$ is a reasonable value for estimates of $W$.

Typically, for $\gamma_{I} = 10$, $\gamma_{p} =30$, values which give an adequate growth rate for the mode (see Jones 2012a,b), with $x = 3.2$, $y = 2$, we have
$W = 6.9$ GeV.  This is of the order of magnitude of the $\epsilon$ values in Table 1.  It is a consequence of the nucleonic masses of the source and the substantial difference between the Lorentz factors $\gamma_{I,p}$.  The value of $W$ varies substantially as a function of $x$ and $y$.  A further possibility is that as radio-frequency emission moves $\gamma_{I,p}$ towards $\gamma_{c}$, further acceleration has the opposite effect and so may obviate the limit given by equation (14).  These factors may possibly contribute to the substantial variations in $\epsilon$ seen in the Table.

\section[]{Multi-component dispersion relation}

Extraction of energy from the longitudinal kinetic energy of the plasma with the efficiency demonstrated by the Table 1 values of $\epsilon$ must presumably rely on growth of an unstable longitudinal mode.  The dispersion relation $\omega(k)$ in the general, quasi-longitudinal case was given by Asseo, Pelletier \& Sol (1990), but for present purposes the longitudinal mode will suffice for which the angular frequency $\omega(k)$ satisfies,
\begin{eqnarray}
\sum_{i}\frac{\omega_{i}^{2}}{\gamma_{i}^{3}(\omega - kv_{i})^{2}} = 1, \hspace{1cm}
\omega_{i}^{2} = \frac{4\pi n_{i}q_{i}^{2}}{m_{i}},
\end{eqnarray}
for a multi-component plasma with $\delta$-function velocity distributions.
Component $i$ has charge $q_{i}$, and $n_{i}$, $v_{i}$ and $\gamma_{i}$ are its number density, velocity and Lorentz factor in the neutron-star frame.

The nature of the plasma depends on the relation between the spin vector and the magnetic dipole moment.  Neutron stars with polar-cap ${\bf \Omega}\cdot{\bf B} > 0$ have at most two components, electrons and positrons.  The energy distributions are not $\delta$-functions as in equation (12) but are more likely to be broad continuous distributions, and further development of an expression to replace equation (15) would be required to calculate $\omega(k)$ and to demonstrate that there is an adequately large growth rate. It would appear at present that no such detailed studies have been published.  Even if the Penrose condition were satisfied, the growth rate does depend on the detail of the velocity distributions.  That this is so is indicated by the solution of equation (15) for its complex pair of roots in the case of electrons and positrons with $\delta$-function Lorentz factors $\gamma_{-}$ and $\gamma_{+} = \gamma_{-} + \delta\gamma$ and for wave-number $k = 2\gamma_{-}^{1/2}\omega_{e}$. This shows that the imaginary part of $\omega$ is of the order of $\omega_{e}\gamma_{-}^{-3/2}(\delta\gamma/\gamma_{-})$.  Therefore, a large growth rate requires well-separated Lorentz factors, as we observed in Section 3.2.

The very basic model of equation (12) is unrealistic, but would feature well-separated Lorentz factors. In this case, evaluation of equation (15) does admit a complex pair of roots for wave-numbers
$k < 2\gamma_{1}^{1/2}\omega_{1}$.  But here the growth rates do not appear to be large and, for small $\gamma_{1}$, mode growth is limited by the possibility of particle momentum reversal in the neutron-star frame.  It is unfortunate that the form of the electron-positron distribution is not known.

The plasma will, in general, have four components in an ${\bf \Omega}\cdot{\bf B} < 0$ neutron star. The mode and its growth rate have been described in previous papers (Jones 2012a,b) in the case where electron-positron number densities are negligible.  But if there were an appreciable pair density from conversion of curvature radiation or from inverse Compton scattering, all four terms may need to be retained in equation (15).  The significant parameter determining the relative contributions of electrons and positrons and of protons and ions is,
\begin{eqnarray}
D = \frac{N_{\pm}m_{p}\gamma_{p}^{3}}{m_{e}\gamma_{e}^{3}}.
\end{eqnarray}
In the general case, it is unfortunately difficult to estimate the extent to which the mass ratio in (16) is likely to be off-set by either of the other factors.
But if prolific pair creation by curvature radiation exists, particularly in the case of high polar-cap magnetic flux densities, $N_{\pm}$ can be large and
$\gamma_{e} = \gamma_{+,-}$ small so that equation (16) gives $D \gg 1$ and the mode angular frequency $\omega$ will be such that ion-proton components are negligible in equation (15). In such cases, the sign of ${\bf \Omega}\cdot{\bf B}$ has no relevance.

The mode angular frequency in the neutron-star frame is $\omega \approx 2\gamma_{i}^{1/2}\omega_{i}$.  This is to be associated with the formation of turbulence as described in Section 3 and therefore with the lowest frequencies observed in the radio-frequency spectrum.  The transfer of energy to higher wave-numbers is rapid and we assume with reference to the properties of other turbulent systems that it leads to the spectra observed with large negative indices.  Thus the whole spectrum is formed within a compact region so that no very obvious radius-to-frequency mapping is expected (see, for example, Hassall et al 2012, 2013).

In general, mode frequencies arising from the two cases differ typically by two or more orders of magnitude reflecting the masses and number densities present in equation (15).  This has been referred to previously in relation to PSR B1133+16 (see Jones 2013b).  To understand the form of radio-frequency spectrum to be expected in a given class of pulsar, we have first to establish the likely nature of the plasma and then make reference to equation (15).  This is attempted for some common classes of pulsar in the following Sections.  The present paper has limitations shared with previous papers (Jones 2012a,b; 2013a,b). It attempts no more than to see the extent to which ion-proton plasma sources are consistent with the broad characteristics of distinct classes of pulsar. Thus the finer details of observational results and problems such as polarization are not addressed.

Coherent curvature radiation was not considered in relation to the black box argument of Section 3 although it is a consequence of the transverse acceleration arising from flux-line curvature and so is unaffected by the kinematic constraints derived there. Significant power-loss in the pulsar context requires
coherence in the form of a macroscopic bunch of co-moving particles. But the in vacuo expression for the power-loss also assumes coherence in the radiation field the bunch produces over a path length of the order of $\gamma^{2}c/\omega$, for neutron-star frame Lorentz factor $\gamma$ and radiation angular frequency $\omega$, and therefore that the charge and spatial structure of the bunches do not change much in that time.  Given these constraints, it may be the case that curvature radiation from the Langmuir-mode induced turbulence contributes to the radio-frequency emission.  But loss of energy by other processes does imply damping of turbulent charge-fluctuations, as we have shown in Section 3.

However, the main point we wish to make is as follows.  This paper assumes throughout that the polar-cap acceleration field ${\bf E}_{\parallel}$ is subject to space-charge-limited flow boundary conditions.  The reason is that extensive calculations have shown that ionic work functions are too small to maintain the ${\bf E}_{\parallel}\neq 0$ boundary condition at a polar-cap surface (see Jones 1985, Medin \& Lai 2006) except possibly at fields of $10^{14}-10^{15}$ G.  We regard fields of this order as unlikely to be universal in a population of neutron stars whose spin-down inferred magnetic dipole moments vary by more than five orders of magnitude.

The basic assumption of the modern form of curvature radiation model (see Gil, Lyubarsky \& Melikidze 2004; Dyks \& Rudak 2013 and references therein) is the
${\bf E}_{\parallel}\neq 0$ condition.  The plasma state required for coherent curvature radiation is then produced by what is referred to as a modulational instability arising from the intermittent spark discharges which are assumed to be a consequence of this boundary condition.  We do not accept this model of the polar cap but also do not regard the present paper as a suitable place for its detailed criticism.  Its purpose is merely to give an alternative view.

\subsection{Normal pulsars}

The selected set of 29 pulsars can be taken to represent pulsars that are normal in respect of period and magnetic flux density.  The parameter
$X = B_{12}P^{-1.6}$, where $B_{12}$ is the magnetic flux density in units of $10^{12}$ G, is a convenient measure of the capacity to form secondary electron-positron pairs in a dipole field, and is drawn from the work of Harding \& Muslimov (2002).  Only B0919+06 and B1933+16 have values exceeding $X_{c} = 6.5$,  the curvature-radiation pair threshold. The remaining pulsars have values of $X$ mostly well below $X_{c}$, and pair formation would require either inverse Compton scattering (ICS) or flux-line curvature larger than that of a dipole field.  Pair formation from ICS photons produces electron-positron energy spectra extending to high energies (Harding, Muslimov \& Zhang 2002) and are unlikely to lead to a mode with a useful growth rate in the ${\bf \Omega}\cdot{\bf B} > 0$ case.  The essence of ICS is that  a large fraction of the momentum of an outward-accelerated electron is transferred to a photon and it follows that it must be much less significant in ${\bf \Omega}\cdot{\bf B} < 0$ pulsars, in which the flux of accelerated positrons is small.

In the case of ${\bf \Omega}\cdot{\bf B} > 0$, equation (15) can refer only to pairs and we can conclude that after the cessation of incoherent photon emission from the outer magnetosphere, the pulsar is likely to be unobservable under normal circumstances because coherent radio emission from a polar-cap source would be problematic for the reasons considered in Section 3.1.

Pulsars with ${\bf \Omega}\cdot{\bf B} < 0$ and with negligible pair densities  therefore constitute the bulk of the population seen in a $P - \dot{P}$ plane distribution.  A particular example, B1133+16, has been discussed previously in some detail (Jones 2013b) where it was shown that the typical observer-frame mode frequency to be expected from an ion-proton plasma was $10 - 60$ MHz, that is, in the turn-over region for many pulsars.

\subsection{Gamma-ray pulsars}

These have been catalogued recently by Abdo et al (2010, 2013).  Our remarks in this Section concern those classified as young or middle-aged and exclude millisecond pulsars (MSP) that are gamma-emitters.  These are further classified into two groups, radio-loud and radio-quiet, of roughly equal number.  The existence of the radio-quiet pulsars may be simply a consequence of differing emission geometries.  All pulsars in the first catalogue (Abdo et al 2010), apart from the 7 MSP, have fields and rotation periods such that $X > X_{c}$. Phase-lags between the radio and gamma-ray peaks vary.  There is a consensus that the source of the gamma emission is close to the light cylinder and it is also possible that the source of radio emission may be located away from the polar cap.

In the ${\bf \Omega}\cdot{\bf B} > 0$ case, curvature radiation pair creation is likely at the polar cap with the formation of a plasma as described in Section 3.1.  Coherent radio emission is subject to the uncertainties we have described above, but if it occurs, the observer-frame frequency at the onset of turbulence would be typically about three orders of magnitude larger than that cited in Section 4.2, that is, more than $10$ GHz, assuming a pair multiplicity $N_{\pm} > 10^{2}$.  This is not consistent with the observed spectra which appear to have indices at $1.4$ GHz, little different from normal pulsars.

Predictions for pulsars with ${\bf \Omega}\cdot{\bf B} < 0$ are less certain.
In this case, the condition $X > X_{c}$ is not sufficient for curvature radiation pair creation.  The reason is that the relatively high surface temperatures to be expected in the younger stars cause the photo-ionization of accelerated ions to a degree not considered in previous work (see Jones 2012a).  The reverse flux of electrons then limits the acceleration potential difference analogously with pair creation.  It is difficult to make useful quantitative estimates of this effect owing to its dependence on surface temperature and on the atomic number of surface nuclei.  Coherent emission, as in normal pulsars, is possible but the system may behave as in the ${\bf \Omega}\cdot{\bf B} > 0$ case. This conclusion has been arrived at previously by Kunzl et al (1998) who found the $160$ MHz emission of the Crab pulsar incompatible with a large-multiplicity pair plasma for any emission source within the light cylinder.

\subsection{Millisecond pulsars}

The conclusion of Jenet et al (1998) was that the radio emission of J0437-4715 differed little from that of normal pulsars and that consequently, the emission mechanism must be insensitive to both magnetic flux density and rotation period.
Further evidence for this, with measurements of spectral indices, came from a study of a sample of 14 millisecond pulsars by Kramer et al (1999), and further work with emphasis on gamma-emitters by Espinoza et al (2013).  Kramer et al found that the emission region is extremely compact (a light-travel distance $< 2.4\times 10^{5}$ cm), but subject to some reservations they make about their interpretation of the data.

Almost all the MSP listed in Table 2 of Kramer et al have ATNF magnetic fields smaller than $10^{9}$ G. Here we argue that in these cases, formation of an electron-positron plasma by single-photon magnetic conversion which is capable of supporting a useful unstable mode is not possible.  The polarization-averaged attenuation coefficient in the classical limit $\chi \ll 1$,  which is certainly appropriate for MSP is,
\begin{eqnarray}
\lambda_{a} = \frac{m_{e}{\rm e}^{2}}{2\hbar^{2}}\frac{B}{B_{c}}
\frac{k_{\perp}}{k}\left(0.377\exp(-4/3\chi)\right),
\end{eqnarray}
in which $\chi = k_{\perp}B/B_{c}$ (Erber 1966).  For $B = 10^{9}$ G, a photon with $k_{\perp} = 3.5\times 10^{3}$ would be required to give $\chi = 0.08$ and an appropriate coefficient of $\lambda_{a} = 10^{-5}$ cm $^{-1}$.  In the typical MSP dipole field, this would require a photon of momentum $k = 10^{4}-10^{5} m_{e}c$  giving electron and positron Lorentz factors so large that the corresponding terms in equation (15) would be negligible.

The conclusion is that the polar-cap formation of an electron-positron plasma supporting a Langmuir mode with a useful growth rate is not possible.  Thus MSP with ${\bf \Omega}\cdot{\bf B} > 0$ would be radio-silent in the absence of emission from the light cylinder radius.  It is not obvious that polar-cap pair creation is a pre-requisite of gamma-ray emission.  MSP with ${\bf \Omega}\cdot{\bf B} > 0$ certainly have a flux of electrons accelerated to the light cylinder.  But the movement of particles to the light cylinder must be more complex than a single flux of uniform charges from the polar cap: neutron stars with ${\bf \Omega}\cdot{\bf B} < 0$ probably also have electron flow to the light cylinder in order to maintain the net electric charge of the star and magnetosphere at some constant level. (This assumes that the return of positive particles from beyond the light cylinder radius to the opposite pole, via the equatorial current sheet, found in numerical modelling of the outer magnetosphere, is incomplete for general values of the magnetic inclination.  We refer to the work of Bai \& Spitkovsky 2010). High-energy electrons or positrons must be present at the light cylinder if there is to be significant incoherent emission, but if our assumption is correct, this condition can be satisfied by pulsars with ion-proton polar-cap plasmas as well as by those with ${\bf \Omega}\cdot{\bf B} > 0$.

\subsection{Anomalous X-ray pulsars}

Polar-cap magnetic flux densities $B > B_{c}$ are the outstanding feature of the AXP and of magnetars in general, and for a review of pair creation in these objects we refer to Thompson (2008).  Pair creation by resonant scattering of black-body photons is so abundant at surface temperatures exceeding $3\times 10^{6}$ K that the total acceleration potential difference at the polar cap can be no more than of the order of $1 - 10$ GeV.  There is also a contribution from Klein-Nishina scattering. Direct photon conversion to positronium is significant at $B > B_{c}$ (see Herold, Ruder \& Wunner 1985; Usov 2002). In this process, a photon moving outward converts (essentially adiabatically) to a positronium bound state when the threshold $k_{\perp} = 2$ is reached. But in the black-body field of the neutron-star, there is no doubt that the positronium would suffer very rapid photo-disintegration. The consequence of these processes is expected to be a very high flux of secondary electrons and positrons whose energy is low relative to those of young normal pulsars.  This conclusion is likely to be independent of the sign of ${\bf \Omega}\cdot{\bf B}$, and implies that the ion-proton terms in equation (15) will be negligible.

Whether or not the electron-positron mode grows at a useful rate is not clear.
Radio emission has been observed from J1810-197 up to very high frequencies ($1.4$ to $144$ GHz; Camilo et al 2007). The spectrum has only a small negative index within this range.  But the pulses exhibit unusually large fluctuations (Serylak et al 2009).  This is broadly the frequency interval that might be expected from an electron-positron plasma.  But this emission may be further evidence that, as we observed in Section 4.3, particle fluxes to the light cylinder may be less simple than a single uniform charge flux from the polar cap.(For indications of a high-frequency turn-up in pulsar spectra, see Kramer et al 1996.) Higher-frequency flux measurements, though difficult owing to atmospheric opacity, would be of interest.  For magnetars in general, the large negative spectral indices of normal pulsars and MSP should not be observed because any ion-proton plasma would have a negligible effect on the plasma dispersion relation.

\section{Conclusions}

Radio-frequency luminosities have been estimated for a homogeneous set of 29 pulsars having extensive low-frequency flux-density measurements and good signal-to-noise ratio.  However, the radio-frequency energy emitted per unit charge in the primary beam of particles accelerated at the polar cap appears to be the more interesting parameter.  Pulsars are very efficient generators of radio-frequency radiation below $1 - 10$ GHz, particularly in view of the kinematic constraints on $\epsilon$ which are discussed in Section 3.  These were considered separately in Sections 3.1 and 3.2 for the canonical electron-positron plasma and for the ion-proton plasma described previously (Jones 2012a,b).

The conclusion of this paper is that, both in normal pulsars and in MSP, an ion-proton plasma is the source of the radio-frequency spectrum and that consequently, these pulsars have a magnetic moment such that ${\bf \Omega}\cdot{\bf B} < 0$ at the polar cap.  A number of factors together form the basis for this conclusion.

(a)  The existence of an ion-proton plasma in the ${\bf \Omega}\cdot{\bf B} < 0$ case is a natural consequence of excitation of the nuclear giant-dipole state in the electromagnetic showers produced by the reverse flux of photo-electrons.  These are produced by the interaction of accelerated ions with the blackbody radiation field of the neutron-star surface.  Ions and protons necessarily have very different Lorentz factors and $\delta$-function velocity distributions.  Conditions are appropriate for a (quasi-longitudinal) Langmuir mode whose growth can be calculated directly.

(b) In this case, the polar-cap magnetic flux density has just one function: to provide an acceleration potential difference. (Throughout this work we assume space-charge-limited flow boundary conditions.)  Photoelectric transitions and the processes of electromagnetic shower development are slowly varying functions of magnetic field strength above $\sim 10^{12}$ G but below this, can be satisfactorily approximated by their zero-field limits.  This is in contrast with the formation of an electron-positron plasma whose production by curvature radiation in a normal pulsar requires a substantial decrease in flux-line radius of curvature from that of a dipole field (see, for example Harding \& Muslimov 2011).  With a dipole field, even the inclusion of inverse Compton scattered photons produces pair multiplicities that are too small to be plausible (Hibschman \& Arons 2001, Harding \& Muslimov 2002). In the case of MSP, polar-cap pair creation is either not possible or results in electrons and positrons of too high an energy to give unstable-mode solutions of equation (15) having a useful growth rate.
The ion-proton plasma appears to be the only mechanism capable of functioning over five orders of magnitude in magnetic flux density.

(c) There are a number of problems which make the electron-positron source unlikely. The relative displacement of the neutron-star frame electron and positron energy spectra is difficult to estimate owing to its dependence on the details of their low-energy extremes.  Our attempts to investigate this in Section 3.1 indicate that the displacement is unlikely to be other than small, leading to severe constraints on values of $\epsilon$ and to doubts about the growth rate of the Langmuir mode.  The form of the electron and positron spectra at low energies is not well-known and it appears that no actual growth-rate calculations have so far been attempted with realistic energy distributions.
These doubts about mode growth rate are not new (see Usov 1987).

(d)  Stringent limits now exist on the size of the emission region, particularly at low frequencies with the operation of LOFAR.  Hassall et al (2012) found a limit defined by $r_{max} - r_{min} < 4.9\times 10^{6}$ cm in the case of PSR B1133+16, with similar limits in the case of three other pulsars. At higher frequencies and for 14 millisecond pulsars, Kramer et al (1999) found $r_{max} - r_{min} < 2.4\times 10^{5}$ cm, though with some stated reservations about the interpretation of their data.  We have argued previously (Jones 2013b) that the Hassall et al data are not consistent with an electron-positron plasma source.

(e)	 There is no direct evidence that neutron stars, other than those young enough to be associated with nebulae, generate electron-positron pairs.

As a consequence of these factors, our conclusion is that, in general, the source of radio-frequency emission is not the canonical polar-cap electron-positron plasma previously assumed.  Electron-positron plasma sources are at best problematic for the reasons we have listed, but we do not wish to exclude them completely.

A further difference from much previous work is in our consideration of the spectrum, of which we relate only the low-frequency end to a local plasma frequency Lorentz-transformed to the neutron-star or observer frame.  In our view, the significant mode frequency is that at which the amplitude has grown to the extent that a non-linear and then turbulent field of charge and current-density fluctuations is formed.  It is then assumed that, as in fluid mechanics (see, for example, Batchelor 1967), there is a rapid transfer of energy to higher wave-numbers.  Thus the mode frequency in the neutron-star or observer frame at the on-set of turbulence is to be associated with the low-frequency end of the radio spectrum and the power-law decrease in flux density reflects the energy transfer to higher wave-numbers.  It is also consistent with the compact emission regions found. Therefore, there is no prediction that the radius-to-frequency mapping, $\nu \propto \eta^{-3/2}$ should be seen. This is consistent with the lack of evidence for it, except for some profile broadening at very low frequencies, in those pulsars whose emission profiles have now been observed over frequency intervals of more than three orders of magnitude in width (see, for example, Hassall et al 2012, 2013).

A consequence of our conclusion is that ${\bf \Omega}\cdot{\bf B} > 0$ neutron stars start as gamma-ray, and possibly radio, pulsars early in their lives and then usually become unobservable.  This does not appear to represent any problem because the log-log $P-\dot{P}$ distribution demonstrates that there is a very considerable loss of population density after $1 - 10$ Myr.

A number of problems have not been addressed here.  Among them are the mechanism of coherent emission in gamma-ray emitters about which Section 4.2 has no clear conclusion, and the peculiar nature of the emission from J1810-197 mentioned in Section 4.4.  There is also the nature of the turbulence assumed.  The coupling between different wave-number components is electromagnetic, and must be fast, but there are differences from homogeneous isotropic fluid turbulence. Firstly, the plasma is clearly not isotropic owing to the presence of a strong magnetic field. But perhaps more serious is the fact that (dissipative) coupling with the radiation field is always present.  There is no dissipation-free interval in which interaction between different wave-number components transfers energy to higher wave-numbers. It remains to be seen if any of these problems are significant.

\section*{Acknowledgments}

The author thanks Dr Maciej Serylak for bringing the case of J1810-197 to his attention.

\bsp

\label{lastpage}

\end{document}